# Interplay between Defects, Disorder and Flexibility in Metal–Organic Frameworks

Thomas D. Bennett,[1]* Anthony K. Cheetham,[1]
Alain H. Fuchs[2] and François-Xavier Coudert[2]*

[1] Department of Materials Science and Metallurgy, University of Cambridge,
27 Charles Babbage Road, Cambridge CB3 0FS, United Kingdom
[2] Chimie ParisTech, PSL Research University, CNRS,
Institut de Recherche de Chimie Paris, 75005 Paris, France.
* e-mail: tdb35@cam.ac.uk; fx.coudert@chimie-paristech.fr

**Metal–organic frameworks are a novel family of chemically diverse materials, which are of interest across engineering, physics, chemistry, biology, and medicine-based disciplines. Since the development of the field in its current form more than two decades ago, priority has been placed on the synthesis of new structures. However, more recently, a clear trend has emerged in shifting the emphasis from material design to exploring the chemical and physical properties of those already known. In particular — while such nanoporous materials were traditionally seen as rigid crystalline structures — there is growing evidence that large-scale flexibility, the presence of defects and long-range disorder, are not the exception, but rather the norm, in metal-organic frameworks. Here we offer some perspective into how these concepts are perhaps inescapably intertwined, highlight recent advances in our understanding, and discuss how a consideration of the interfaces between them may lead to enhancements of the materials' functionalities.**

As an entirely novel class of nanoporous materials, metal–organic frameworks (MOFs) have attracted a lot of interest and are studied by scientists from a wide variety of horizons, due to both their intrinsic aesthetic appeal and their potential industrial applications, for example as selective adsorbents and catalysts, substrates for biosensors and drug delivery, and membranes and films in various nanotechnologies, to list but a few. Compared with both dense and nanoporous inorganic materials, MOFs are based on relatively weaker interactions (coordinative bonds, π−π stacking, hydrogen bonds, etc.) and present large numbers of intramolecular degrees of freedom. While all molecular assemblies and solids show some degree of flexibility, even those considered "rigid", like the archetypical MOF-5, there is a propensity among coordination polymers to display large-scale dynamic behaviour, which is typically described by the vague term "flexibility"[1] (Fig. 1). This appellation contains phenomena that are very diverse both in terms of their microscopic origins and their macroscopic manifestations. The latter include negative thermal expansion, a relatively common trait[2], as well as anomalous mechanical properties[3], such as auxeticity (the presence of a negative Poisson's ratio, where a material becomes thicker when stretched along certain directions) or negative linear compressibility (total volume reduction with expansion along a specific direction under mechanical pressure)[4]. Both phenomena are rare among inorganic and molecular solids and stem from wine rack, honeycomb and related topologies, which allow the framework to act as a Tinkertoy model with rigid struts and hinges[5].

Just as there is flexibility in any molecular assembly, there is no such thing as a crystalline solid without defects. The prevalence of defects in nanoporous frameworks was first made evident by the difficulty in comparing fluid adsorption isotherms across experimental measurements performed by different groups, and against computational predictions. The most commonly observed effect was an incomplete activation of framework porosity, resulting in guest molecules blocking the nanopores and negatively impacting surface area and adsorption capacity[6,7]. It was only after some time that defects intrinsic to the frameworks, for example linker and inorganic node vacancies, partial metal reduction and dislocations, started to be characterized and reported[8]. Recently, Li and co-workers described how defects may be introduced into the formation mechanism of some MOFs, leading to mismatched growth and gelation, as opposed to crystallisation in long-range ordered structures[9]. Indeed, it is now being recognised that such defects need not necessarily have adverse effects but can give rise to specific functionalities, such as improving adsorption affinity or catalytic activity[10].

Hoskins and Robson's description of MOFs in 1990 as 'ordered, truly crystalline' structures is recognised as a milestone in the field.[11] Subsequently, examples of framework dynamics or rotational disordering of cations within MOFs, along with cases which consider topologically or statically disordered materials (that is, X-ray amorphous species possessing no long range order), perhaps challenge this view[12,13]. Comparatively few reports of such systems exist, though it is likely that many have been synthesized and subsequently discarded due to their lack of crystallinity. Interestingly however, even Robson and Hoskins recognised that their approach to MOFs could yield 'amorphous materials… with many (internal) loose ends'[11]. Indeed, recent terminology guidelines[14] now do not necessarily limit MOF materials to those which exhibit long range ordering, and disordered MOFs are becoming of increasing interest to the community, driven by an emerging focus on the *physical* properties of framework structures.

Flexibility, defects and disorder are all related to entropy (Fig. 1). They arise from the high dimensionality of the intramolecular degrees of freedom of the materials — and correspondingly many "soft" modes of low energy — as well as the relatively weak nature of the interactions involved in the framework assembly, such as coordinative bonds, $\pi-\pi$ stacking and hydrogen bonds. For example, the MIL-53(Al) structure undergoes a thermally-induced phase transition in the absence of solvent between 325 and 375 K. The low temperature narrow pore form is stabilized by dispersion interactions, though it is the contribution of vibrational entropy that drives the formation of the more porous structure at high temperatures[15]. Entropy also plays an important role in the formation of disordered MOF systems upon heating[16], and, in some MOFs, proves sufficient thermodynamic driving force to overcome unfavourable enthalpies of defect incorporation[8].

## *Flexibility and Defects*

One aspect of the flexibility of MOFs is their general softness – in the terminology of elasticity – or compliance, i.e. their high compressibility compared to traditional nanoporous materials. This has been widely observed and discussed in the literature[17], based on both experimental measurements and quantum chemistry predictions, and is linked to the weaker nature of their bonds (compared, for example, to the very strong Si–O bond in zeolites), and to their higher porosities. More dense MOFs or those with very high connectivities feature the highest elastic moduli, with bulk moduli in the 25–50 GPa range[18]. Flexibility in MOFs is, however, tightly coupled with the potential presence of defects, as can be illustrated in the UiO-66 family of porous isoreticular frameworks.

With a chemical formulae based on $[M_6O_4(OH)_4][C_6H_4(COO)_2]_6$ (M = Ti, Zr, Hf, …) and consisting of 12-connected nodes, these materials display relatively low compressibility as a result. UiO-66 not only presents substantial non-stoichiometry, with up to 10% of terephthalate linkers missing, but the concentration of defects can be controlled in a systematic manner by varying the concentration of a modulator (typically a monocarboxylic acid) during the synthesis[19]. The capacity of the framework to accommodate a large number of missing linkers has a strong impact on the materials' properties, both by changing the characteristics of the pore space of the MOF and by introducing sites with higher catalytic activities. In UiO-66(Zr), the inclusion of linker vacancies was used to tune the pore volumes from 0.44 to 1.0 cm$^3$ g$^{-1}$ and the specific surface from 1,000 to 1,600 m$^2$ g$^{-1}$ [20], with a gas adsorption performance for $CO_2$ that is higher than the "perfect" defect-free crystal. Vermoortele *et al*. showed that post-synthetic removal of the modulators can lead to large concentrations of coordinatively unsaturated (*cus*) metal sites, and drastically increased activity for several Lewis acid catalyzed reactions[21].

However, the missing linkers also affect the mechanical properties and flexibility of UiO-66 (Fig. 2). Decreasing the number of bidentate linkers from an average of 12 (in the 'defect-free' structure) to 8, was observed to result in a decrease in bulk, elastic and shear moduli. Anisotropy was also observed to increase upon increasing defect concentrations, whilst, curiously, auxeticity was also introduced at these higher defect concentrations[22]. The effect on $CO_2$ uptake of such defects was also studied, with high-pressures eliciting increased capacities in the defective structures.

Another example of the intimate interplay between flexibility and defects is based on the use of linker substitution (as opposed to linker removal) to create multivariate MOFs, i.e. frameworks encompassing several distinct functionalities by the incorporation of different chemical groups on their organic linkers[23,24], through mixing functionalized linkers based on the same backbone[25,26]. A series of mixed ligand compounds, starting from the interdigitated layered structures of {Zn(5-NO$_2$-ip)(bpy)}(0.5DMF·0.5MeOH) and {Zn(5-MeO-ip)(bpy)}(0.5DMF·0.5MeOH) were prepared by Kitagawa *et al*. (ip = isophthalate, bpy = 4,4′-bipyridyl). The flexible *gate-opening* behaviour (transition between a closed form and an open form) of the former compound upon $H_2O$ and $CO_2$ sorption was found to be substantially modified through solid solution formation (Fig. 2). Selective adsorption uptakes of $CO_2$ from a $CO_2/CH_4$ mixture were investigated, revealing selectivities in the mixed linker frameworks that overcame the low overall adsorption of the pure 5-$NO_2$-ip linker system, and the poor separation abilities of the pure 5-MeO-ip linker containing system.

Post-synthetic modification offers perhaps a more systematic approach to linker substitution, such as the use of solvent-assisted linker exchange (SALE) in Zn- and Cd-based zeolitic imidazolate frameworks (ZIFs)[27,28]. Here, Hupp and co-workers demonstrated that by exchanging linkers in a relatively simple framework, one can introduce other chemical functions in a manner that is controlled by the exchange kinetics. This process can even be used to synthesize frameworks that were previously unobtainable by direct synthesis methods, because they would crystalize in another – undesirable – topology. Several examples of the use of SALE in flexible, low-density ZIF structures exist[29], and the resulting materials show a broad variety of thermal and mechanical behaviour and stability[30].

### *Flexibility and Disorder*

There are several examples of temperature-induced disorder being used to create flexible materials, showing how entropy intertwines flexibility and disorder in MOFs. Resonant ultrasound spectroscopy was used to probe phase transitions in the formate MOF of composition $[NH_4][Zn(HCOO)_3]$. It was found that the low temperature ferroelectric structure, which adopts the polar hexagonal space group $P6_3$, underwent a transition to a paraelectric phase upon heating over 192 K. The high temperature framework exists in the non-polar hexagonal space group $P6_322$, and the transition is associated with significant elastic softening due to dynamical disordering of the $NH_4$ cations within framework channels (Fig. 3)[31].

A further example of engineering disorder in systems and using it to create flexibility can be found in the UiO framework family. In UiO-67, a planar biphenyl carboxylate ligand connects two $Zr_6O_4(OH)_4$ nodes, and molecular dynamics simulations show good agreement between simulated and crystallographic structures[32]. However, replacement of the linker with azobenzene-4,4′-dicarboxylate, in which the two central nitrogen atoms possess occupational disorder, causes significant dynamic disorder in which the ligands bow in and out of the plane. Consequently, the azobenzene derivative possesses significantly greater flexibility, which is reflected in a significantly lower elastic modulus.

On the other hand, flexibility can also be used to induce disorder. MOFs possess limited mechanical stability, which results in low resistances to structural collapse, or amorphization. Such instabilities have been evidenced under various conditions: hydrostatic compression – both with nonpenetrating and penetrating pressure-transmitting fluids, uniaxial compression, and ball milling. The microscopic picture of this mechanical instability, established through quantum chemistry calculations as well as experimental phonon measurements, shows that pressure-induced amorphization is linked to the existence of deformation modes of low elastic modulus and their pressure-induced softening (Fig. 3)[30]. This mechanism has now been confirmed on several MOFs of varying chemical composition and topology. In most cases, the pressure-sensitive soft modes are shear modes, which explain why some MOFs can withstand significant hydrostatic pressure yet are unstable upon solvent evacuation (i.e. 1st generation MOFs)[33]. This key finding was first established experimentally on ZIF-8 $[Zn(C_4H_5N_2)_2]$, whose shear modulus (slightly below 1 GPa) is one order of magnitude lower than the bulk modulus (8 GPa)[34], but it is now understood to be a common feature in MOFs.

### *Defects and Disorder*

The highly ordered structures of MOFs are often ascribed to their coordination-based self-assembly and highly directional bonding[35]. For example, Su *et al.* demonstrated that formation of the MIL-53 (Al) framework [Al(OH)(BDC)] proceeded via initial assembly of reactants into MOF nanoparticles of *ca.* 10 nm diameter, before rapid crystallization kinetics led to subsequent aggregation and continued coordination of these clusters into extended MOF particles of *ca.* 200 nm in size[9]. Disruption of this secondary long-range aggregate forming step was observed to cause defective coordination in the end product and yielded disordered particle aggregates, or metal-organic gels (MOGs) (Fig. 4). In the example of $[Cu_2(ndc)_2(dabco)]$, defects were introduced in the aggregation stage by using synthetic modulators such as acetic acid, to provide competing, connectivity-terminating interactions[36]. $Fe^{3+}$ and $Cr^{3+}$ dicarboxylate aerogels were also formed by introducing defects at a key stage in the formation process, and described by James *et al* as a 'coherent, spongy network of continuous particles'[37]. Such work provides pathways to low density, disordered aggregates of MOF particles and the associated inclusion of hierarchical porosity[38].

Instances of defect-induced disorder can also be located within the melting behaviour associated with crystalline MOFs or coordination polymers[12,16,39-41]. In one example, heating a ZIF of composition $[Zn(C_3H_3N_2)_2]$ to *ca.* 550 °C resulted in formation of a macroscopic liquid state. Cooling of the resultant liquid resulted in a glass of identical composition, composed of $ZnN_4$ tetrahedra linked by imidazolate ligands (i.e. the same bonding motif as in the starting crystalline state)[16]. A mixed ligand variant was subsequently prepared by addition of dopant quantities of a secondary linker during the initial synthesis, and the resultant, topologically identical material (ZIF-62, $[Zn(Im)_{1.75}(bIm)_{0.25}]$) was observed to melt at a strikingly lower temperature, of *ca.* 400 °C[16].

Linker substitution is not the only example of defects being used to influence disorder in hybrid glasses. The addition of network modifier ions (i.e. creation of network defects) to lower the melting point of silicates is an established technique in the glass domain[42]. A related route has already been employed by Kato and Kasuga[43] in the pursuit of proton-conductive hybrid glasses. Here, taking advantage of the known transfer of protons between imidazolate-type ions[44] a zinc phosphate glass was heated with a benzimidazolate additive to 160 °C, resulting in the formation of a hybrid, proton-conducting material. The building in of defects such as lanthanide or transition metal ions into crystalline structures prior to melting, and subsequently carrying it through into the glass phase, would appear to be a viable and attractive route to functional disordered materials.

Given the propensity for defects in other MOF families, the strategy of utilizing defects to induce melting behaviour may hence be applicable to other MOF systems, aside from ZIFs. Alternatively, the hot-pressing method utilized by Horike *et al.* to form a glass from $[Cd(H_2PO_4)_2(1,2,4\text{-triazole})_2]$ could also be employed for glass formation[45].

## *Connecting Flexibility, Defects and Disorder*

The occurrence of flexibility, defects and disorder in MOFs are inevitably interconnected, though the intricacies of the relationship are only starting to emerge and all aspects are not yet fully understood. Whilst the chemical advantages of combining inorganic and organic components in one system are well established, the physical consequences of this duality are often believed to be detrimental. However, this unique fusion allows flexibilities and disorder, more commonly associated with organic polymers, to be linked with defects reminiscent of classical oxide chemistry. We illustrate here some examples of the complexity arising from the presence of flexibility, defects and disorder.

While bimetallic MOFs have been reported early in the advancement of MOF research, more complex heterometallic MOFs containing larger numbers of cations are now starting to appear, such as Yaghi's family of mixed-metal MOF-74, containing up to 10 different divalent metal cations (Mg, Ca, Sr, Ba, Mn, Fe, Co, Ni, Zn, and Cd) incorporated in their structure[46]. These heterometallic MOFs can have an impact on the performance of the material due to the addition of the functions of metal-based functionalities or through synergistic effects of the heterometals. There is likely a strong effect of the presence of ordering or disorder in the metals' spatial distribution, although this has been little studied so far.

Increases in performance can also be elicited from homometallic materials. For example, MIL-47(V) can exhibit remarkably different breathing behaviours (i.e. unit cell volume change), depending on the oxidation state of the metal center. The structure, in which terephthalate ligands bridge corner sharing $VO_6$ octahedra to yield interconnected chains, was previously demonstrated to exist in a rigid open form, regardless of solvent or guest content. This is very different to other related structures, e.g. MIL-53, where adsorption of solvent elicits a shrinking of the unit cell[47]. Careful evacuation of the as-synthesized MIL-47 material was however observed to prevent oxidation of $V^{III}$ centers to $V^{IV}$, yielding an evacuated material of composition $V^{III}(OH)(bdc)$. Breathing behaviour in this material was observed due to solvent interactions with OH groups on metal centers. An interesting extension on the work was the preparation of mixed valence MIL-47 derivatives by partial oxidation of $V^{III}(OH)(bdc)$. Conductivity measurements performed upon samples showed that whilst single oxidation state derivatives were very poorly conducting, mixed oxidation state samples demonstrated appreciably greater conductivities. In this case, differences were ascribed to the homogeneous distribution of oxidation site defects throughout the material, though the possibility of domain formation was not investigated further[47].

There is now a growing realization that, in addition to ligand and metal substitutions, other types of defects can exist in MOFs, and in particular UiO-66 (Hf). Metal cluster vacancies in the material were shown to be present by Goodwin *et al.*, and controlled by altering the quantity of synthetic modulator. However, it was also demonstrated that the structural disorder introduced by such vacancies was correlated, with defective sites effectively dispersed unevenly into nanoregions of 7 nm size, within a non-defective framework matrix[48]. The existence of these correlated defects had a significant impact upon the elastic, shear and bulk moduli of the material, with decreases of *ca*. 50 % noted in each case. Such increases in flexibility raise the prospect of tuning the materials' performance in charge transport, catalysis, mechanical and optical responses, and will necessitate the development of tools – both experimental and theoretical – to investigate the correlations between defects and their spatial distribution across the MOF family.

These phenomena of defects, disorder and flexibility in MOFs present significant challenges. It has often been the case that they are systematically compared to other porous inorganic or organic systems. A hitherto prevailing thought has been that their principal advantages (chemical versatility and pore size/tunability) are severely tempered by structural fragility, which manifests itself in flexibility, defects and disorder. A wealth of studies is now emerging which looks to seize upon these assumed weaknesses and utilize them for new applications and directions in the field, rather than try and ameliorate or 'solve' the structural fragilities that are now recognised as quintessentially characteristic of MOFs. It is worth noting that a very similar approach was pioneered by de Gennes, in coining the term fragile materials, and thus launching a new field of exploration into soft polymeric matter. Areas where these design principles might be expected to play a particularly beneficial role in the future include proton and ion conductors[49], electrical semiconductors[50] and tuneable optical systems.

**Acknowledgements**
We acknowledge CECAM and CNRS for funding the Workshop on Flexibility and Disorder in Metal–Organic Frameworks (Paris, June 3–5 2015), which spurred discussion of these topics. TDB acknowledges Trinity Hall (University of Cambridge) for funding.

**Author contributions**
All authors initiated this discussion and designed the perspective, T.D.B. and F.-X.C. wrote the manuscript, all authors revised it.

**Additional information**
Correspondence and requests for materials should be addressed to T.D.B. and F.-X.C.

**Competing financial interests**
The authors declare no competing financial interests.

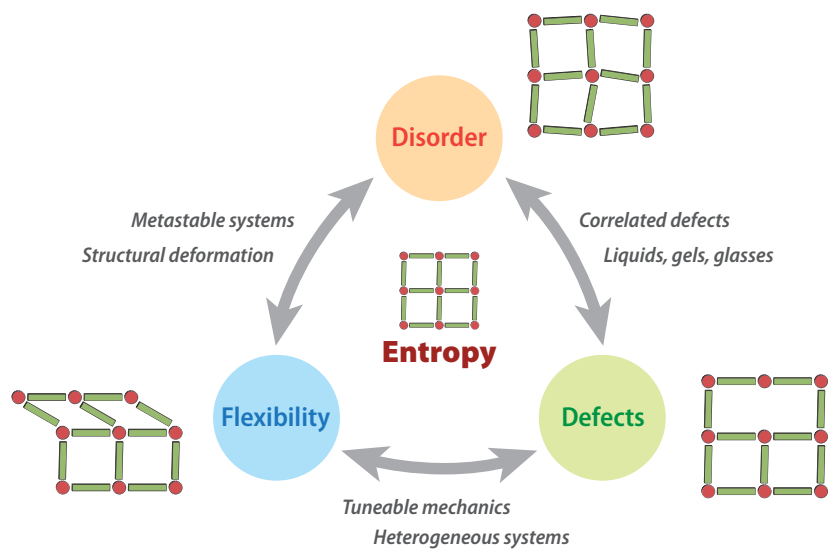

**Figure 1. Flexibility, defects and disorder in metal–organic frameworks.**
Illustration of the interplay between the intertwined concepts of flexibility, defects and disorder in metal–organic frameworks, highlighting some of the phenomena (grey arrows) that emerge from their coupling. Entropy plays a central role in all of these systems.

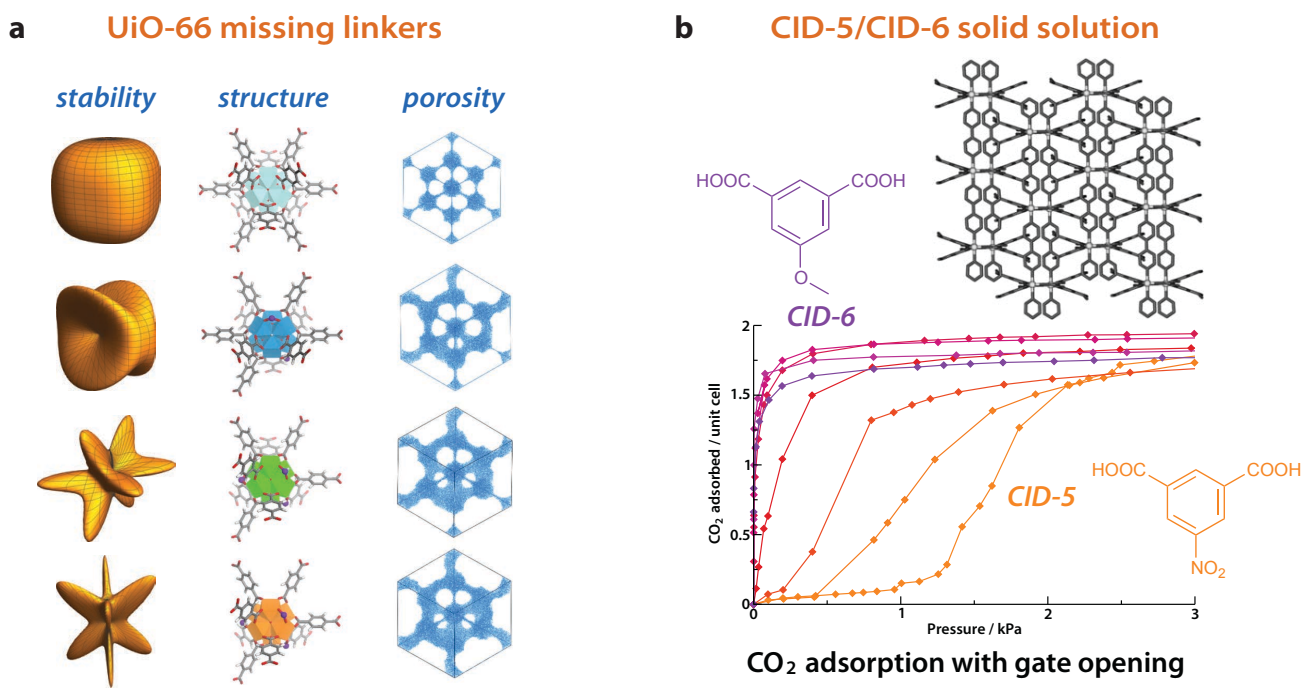

**Figure 2: Interplay between defects and flexibility in metal–organic frameworks.**
**a**, Defect engineering in the UiO-66(Zr) family of metal–organic frameworks is illustrated by plotting the mechanical stability (represented here by directional Young's modulus; leftmost column, in orange) and accessible porosity (rightmost column, in blue) of several defective structures with missing increasing number (from top to bottom) of missing linkers. Adapted from Ref. 22 with permission from The Royal Society of Chemistry.
**b**, Tuning of the gate opening structural transition upon $CO_2$ adsorption in a CID-5/CID-6 solid solution. The carbon dioxide adsorption isotherms at 195 K are plotted for CID-5 (orange), CID-6 (purple), and their solid solutions (intermediate colors). The interdigitated structure of the solid solution is shown in inset (with 0.48% of CID-5 linker and 52% CID-6 linker). Adapted with permission from Ref. 24. Copyright 2010. John Wiley and Sons.

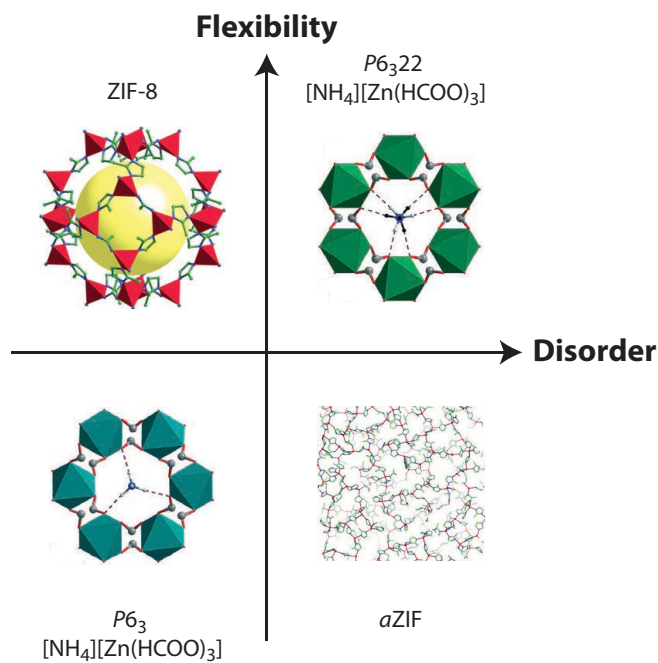

**Figure 3. The relationships between flexibility and disorder in metal–organic frameworks.**
In green circles: crystalline zeolitic imidazolate framework ZIF-8 (top left), which exhibits local flexibility of its linkers and shear instabilities that lead to amorphization upon application of pressure, forming an amorphous ZIF (*a*ZIF, bottom right), which has a disordered network and no local flexibility. In blue circles: the ordered polar (bottom left) and disordered non-polar (top right) systems of a zinc formate framework, $[NH_4][Zn(HCOO)_3]$, whose ferroelectric–paraelectric transition is associated with significant elastic softening due to dynamical disordering of the $NH_4^+$ cations within framework channels. Figure adapted from Ref. 31 by permission of The Royal Society of Chemistry.

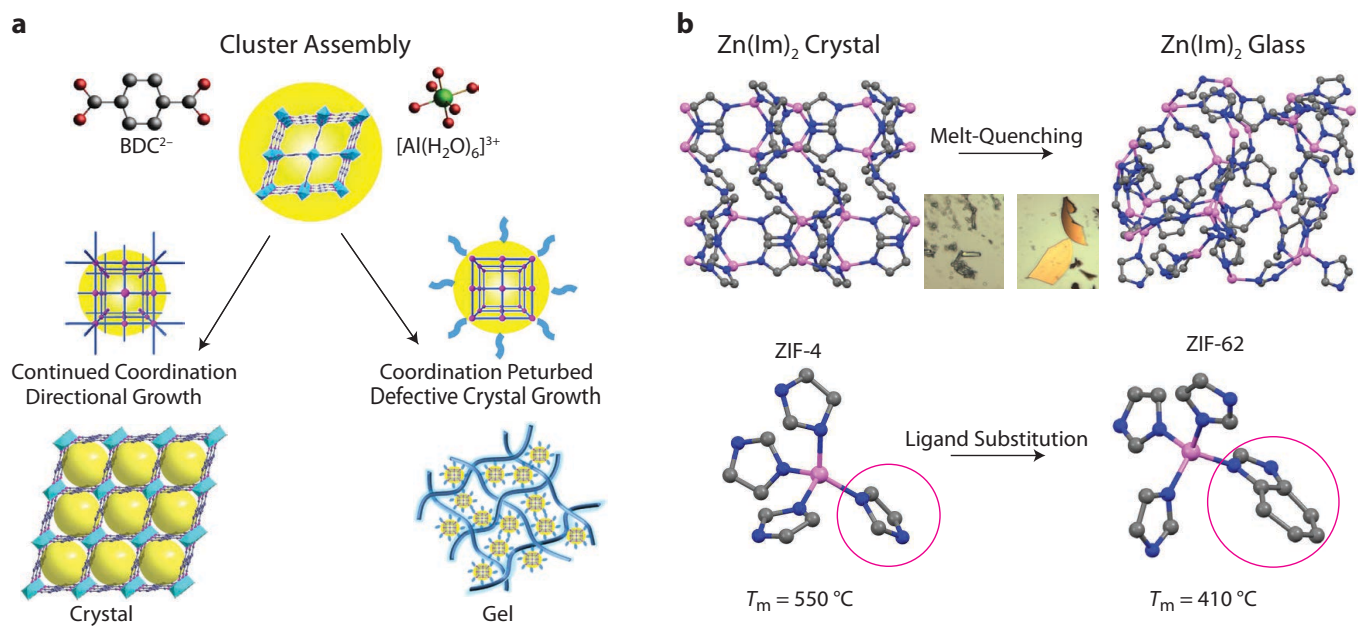

**Figure 4. Coupling between defects and disorder in metal–organic frameworks.**
**a**, Introduction of mechanistic defects, by perturbing coordination during the formation of MIL-53(Al), leads to the formation of a gel based on MOF nanoparticles — instead of the crystalline product obtained by unperturbed synthesis. Removal of the solvents via subcritical carbon dioxide extraction then yields a porous metal–organic aerogel with hierarchical porosity. Adapted from Ref. 9 with permission from Macmillan Publishers Ltd.
**b**, Defect introduction in zeolitic imidazolate frameworks (ZIFs) leads to a lowering in melting point. Partial substitution of the imidazolate linker of ZIF-4 by benzimidazolate (forming ZIF-62) lowers the melting point of the material from 550°C to 410°C, allowing to form novel MOF glasses by melt–quenching. Adapted from Ref. 16 with permission from American Chemical Society.